\documentclass[twocolumn,aps,pra,showpacs]{revtex4}

\usepackage{graphicx} % Include figure files
\usepackage{dcolumn} % Align table columns on decimal point
\usepackage{bm} % bold math
\usepackage{amsmath}
\usepackage{amssymb}

\begin{document}

\title{Optical repumping of triplet $P$-states enhances magneto-optical trapping of ytterbium atoms}

\author{Jun Woo Cho$^{1,2}$}
\author{Han-gyeol Lee$^{1}$}
\author{Sangkyung Lee$^{1}$}
\author{Jaewook Ahn$^{1}$}  \email{jwahn@kaist.ac.kr}
\author{Won-Kyu Lee$^{2}$}
\author{Dai-Hyuk Yu$^{2}$}
\author{Sun Kyung Lee$^{2}$}
\author{Chang Yong Park$^{2}$} \email{cypark@kriss.re.kr}

\affiliation{$^{1}$Department of Physics, KAIST, Daejeon 305-701,
Korea}
\affiliation{$^2$Korea Research Institute of Science and
Standards, Daejeon 305-340, Korea}

\date{\today}

\begin{abstract}
Radiative decay from the excited $^1P_1$ state to metastable
$^3P_2$ and $^3P_0$ states is expected to limit attainable trapped
atomic population in a magneto-optic trap of ytterbium (Yb)
atoms. In experiments we have carried out with optical repumping
of $^3P_{0,2}$ states to $^3P_1$, we observe enhancement of
trapped atoms yield in the excited $^1P_1$ state. The individual decay rate to each metastable state 
is measured and the results show
an excellent agreement with the theoretical values.
\end{abstract}

\maketitle

\section{Introductions}

Ytterbium (Yb, $Z=70$) is a rare-earth element of versatile internal level structure,
generally referred to as singlet-triplet atomic system,
and its narrow intercombination transitions between singlet and triplet spin manifolds
have opened the possibility of many
fundamental studies and applications, including the optical
frequency standards~\cite{Poli,Kohno,Lemke}, parity nonconservation
tests~\cite{Demille}, and ultracold collision and scattering
characterizations~\cite{Weiner, Dinneen, Zinner, Ciurylo, Dzuba,
Machholm}. The natural abundance in Yb isotopes,
%($^{168}$Yb, $^{170}$Yb, $^{171}$Yb, $^{172}$Yb, $^{173}$Yb, $^{174}$Yb, $^{176}$Yb),
furthermore, allows the photo-induced cold molecular
formation of bosonic and fermionic, and composite
dimers~\cite{Kitagawa}.
Recent studies report the sympathetic
cooling of $^{176}$Yb below the transition temperature and the
far-off-resonance trapping of fermionic degenerate $^{173}$Yb as
well as the realization of Bose-Einstein condensation of $^{174}$Yb
and $^{170}$Yb atoms~\cite{Takasu, Fukuhara, Fukuhara2}.

\begin{figure}[tb]
\centering
\includegraphics[width=0.45\textwidth]{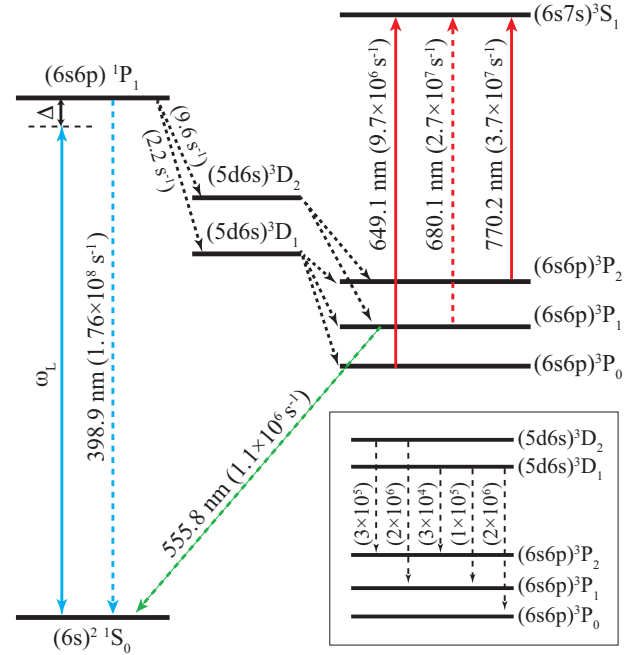}
\caption{Simplified energy level diagram for $^{174}$Yb showing the main cooling transition at 398.9~nm,
radiative channels for the $^1P_1$ excited state, and the relevant repumping scheme (649.1~nm and 770.2~nm).
$\omega_L$  and $\Delta$  are the $^1S_0$-$^1P_1$ cooling laser frequency  and detuning, respectively.
Numbers in parenthesis give the transition Einstein A coefficients.
} \label{fig1}
\end{figure}

The energy level structure and the decay rates of $^{174}$Yb are shown in Fig.~\ref{fig1}.
The blue transition with a relatively broad linewidth allows the optical excitation from the 
($6s^2$)$^1S_0$ ground state  to the ($6s6p$)$^1P_1$ excited state  to be used in both 
Zeeman slowing and magneto-optical trapping~(MOT). The radiative decay from the $^1P_1$ excited 
state  to the $^3P_{0,2}$ triplet states  via the $^3D_2$
and $^3D_1$ states, however, causes the loss of the trapped atoms and, as a result, limits the number of trapped atoms.

In this paper,  we report an experimental demonstration of an $^{174}$Yb
MOT with optical repumping of the metastable states. We use
additional laser systems to control the shelving losses to
${^3}P_0$ and ${^3}P_2$, (i.e. optical repumping of the
metastable atoms to $(6s7s){^3}S_1$ state) and the atoms in the metastable states
continuously decay to the ground state. By doing so, the atom
trapping is only hindered by the collisional loss and, as a
result, the number and lifetime of trapped Yb atoms are increased.
It is shown, in a master equation calculation performed to verify the atom
trapping dynamics, that the trapped-atom collisions
against ballistic Yb flux from the Zeeman slower plays a crucial
role in reducing the measurement uncertainty. 
We have measured the individual decay rate to each metastable state, by minimizing the measurement uncertainty
coming from Yb-Yb and Yb-background gas collisional losses, and the measured results  show
an excellent agreement with the theoretical values.

The paper is organized as following: In Sec.~II,  we describe the
experimental setup where the repumping lasers are used to enhance
the excited atomic population in the MOT.  In Sec III, the
calculated population variations of each levels are given after
solving the master equations. We present the
experimental measurement of decay rate to individual metastable state 
in Sec. IV before concluding in Sec. V.

\section{Experimental Description}
Figure 2 shows the schematic setup diagram for the cooling and
trapping of $^{174}$Yb atoms. A standard six-beam MOT was constructed with
a diode laser system that was injection-locked to the  wavelength 398.9~nm
for the $^{1}S_{0}$-$^{1}P_{1}$ transition ($\lambda_L$ = 398.9~nm,
$\Gamma_{0}$ = 28~MHz (FWHM))~\cite{Park}. The master laser was a home-made
ECDL (External Cavity Diode Laser), that is 15~MHz red-detuned from the $^{174}$Yb
$^{1}P_{1}$-$^{1}S_{0}$ fluorescence peak. Then the master laser
seeded two slave lasers respectively used for the trapping and
Zeeman slowing. The output power of the trap laser was up to 40~mW and, with this laser power,
the fraction of the number of atoms in the excited state, $f$, was varied from 0.05 to 0.2. 
Here $f$ is defined as
\begin{equation}
f=\frac{I/2I_s}{[1+I/I_s+(2\Delta/\Gamma_0)^{2}]}\;,
\end{equation}
where  $I$ is the laser intensity and the saturation intensity is $I_{s} =
59$~mW/cm$^{2}$ for the $^{1}{S}_{0}$-$^{1}{P}_{1}$ transition.

\begin{figure}[tb]
\centering
\includegraphics[width=0.45\textwidth]{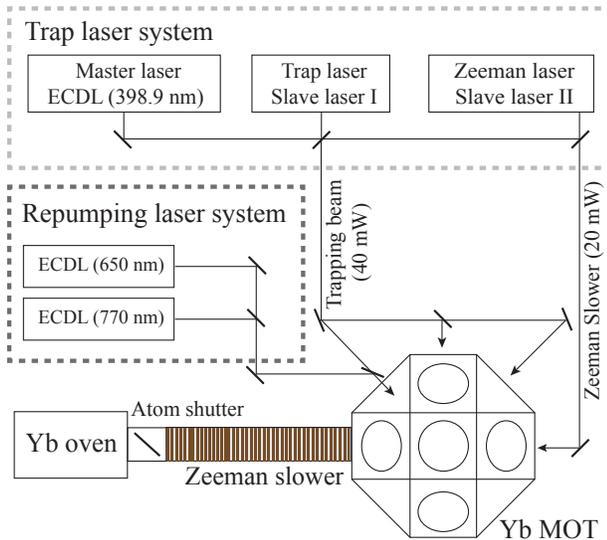} \label{setup}
\caption{ Schematic representation of the setup for the magneto-optical trapping
of $^{174}$Yb atoms with repumping laser systems for  $^3P_2$-$^3S_1$
and $^3P_0$-$^3S_1$ transitions. ECDL is an acronym for external cavity diode laser. }
\end{figure}

For the Zeeman slower laser, a double-pass acousto-optic modulator
was used to further red-detune the frequency by 500~MHz from the
master laser frequency. The output power of the  Zeeman slower
laser was 20~mW. The effusive atomic $^{174}$Yb beam was generated from an
oven heated at a temperature of $400^{\circ}{\rm C}$, while the
oven nozzle was differentially heated at a higher temperature of
$415^{\circ}\rm{C}$. The coil for the Zeeman slower was made of a
1-mm-diameter copper wire, and was wound on a step-like 30-cm-long
stainless-steel pipe with a 16-mm inner diameter. 
When the Zeeman slower laser was operated at the saturation laser intensity, the capture velocity was 260~m/s 
in the Zeeman slower with a peak magnetic field of 300~G. 
In the trap region the axial magnetic filed gradient  $dB/dz=30$~G/cm
made by anti-helmholtz coils.
The Zeeman slower region was designed to maintain nearly
constant deceleration of 80~km/s$^2$ and the atomic beam flux was $10^{10- 11}$~s$^{-1}$.

To investigate the enhancement of the trapped atom yield by
controlling the shelving loss to the triplet states,  the $^3P_0$
and  $^3P_2$ states were repumped to $^3S_1$ by additional lasers.
We used two repumping laser systems (ECDL) at wavelength of
649.1~nm and 770.2~nm lasers, respectively, frequency-locked to  $^3P_0$-$^3S_1$
and $^3P_2$-$^3S_1$ transition fluorescence signals.
For laser frequency calibration, two more laser systems
(ECDL) at wavelength 556~nm and 680~nm were used to induce
two-photon transition $^1$S$_0$-$^3P_1$-$^3S_1$ and the
fluorescence signal was modulation-locked to the $^3P_2$-$^3S_1$
and $^3P_0$-$^3S_1$ transitions.

\section{Rate equations for trap and repumping}

Considering the slow decays from the $^{1}{P}_{1}$ state to the
$^{2}{D}_{1,2}$ compared with ones from the $^{3}{D}_{1,2}$ to the
$^{3}{P}_{0,1,2}$, in the time scale of cooling and trapping
process, an equivalent system for the $^{174}$Yb atom is a 5-level
system. For the states denoted by $|g\rangle$=$^{1}S_{0}$,
$|e\rangle$=$^{1}P_{1}$, $|0\rangle$=$^{3}P_{0}$,
$|1\rangle$=$^{3}P_{1}$, and $|2\rangle$=$^{3}P_{2}$, the rate
equations are given as
\begin{eqnarray}
\nonumber
\frac{d}{dt}({N}_{g}+{N}_{e}) &=& \eta - (a_0+a_1+a_2) N_{e}  \\
&+&  \gamma_{1}N_{1} - \gamma_c (N_{g}+N_{e}), \label{eq2} \\
\frac{d{N}_{0}}{dt} &=& a_{0}N_{e}-\gamma_c N_{0}, \label{eq3}\\
\frac{d{N}_{1}}{dt} &=& a_{1}N_{e}-\gamma_c N_{1} - \gamma_{1} N_{1}, \label{eq4}\\
\frac{d{N}_{2}}{dt} &=& a_{2}N_{e}-\gamma_c N_{2}, \label{eq5}
\label{equation1}
\end{eqnarray}
where $\eta$ is a loading rate of MOT, $a_{{i}
(=0,1,2)}$ are the decay rates from the excited state to the $^{3}P_{i}$ states,
$\gamma_c$ is background collisional loss rate, and $\gamma_1$ is
the spontaneous decay rate of $^{1}P_{1}$. Including the Yb-Yb
collisional term, which is not negligible for some cases in our experiment, 
the equation for the number of trapped atoms $N=N_{g}+N_{e}$ becomes
\begin{eqnarray}
\frac{dN}{dt}=\eta-[a_{2,0} f(P_T, \Delta) + \gamma_c(f)]  N -
\beta(f)N^2\;, \label{eq1}
\end{eqnarray}
where $a_{2,0}$ is the decay rate of
${^1}P_1$ atoms to the both metastable ${^3}$P$_{0,2}$ states and
$\beta(f)$ is the Yb-Yb collision coefficient~\cite{Loftus}. 

In theory, the decay rates are given as $a_{0}= 6.18$~s$^{-1}$,
$a_1=5.25$, and $a_{2}=0.37$~\cite{Porsev}. 
When the atoms in either the $^3P_{0}$ state or the $^3P_{2}$ state 
are optically pumped to the
$^3S_1$ state, they spontaneously decay to the three $^3P_{0,1,2}$
states with a branching ratio of $\lambda_0:\lambda_1:\lambda_2=1:3:5$,
and the atoms in
the $^3P_{1}$ state immediately decays to the ground state. So, we can
consider the following four different repumping cases: (NR) no
repumping case, (A) repumping the $^{3}P_{2}$ state only, (B)
repumping the $^{3}P_{0}$ state only, and (A+B) repumping both the
$^{3}$P$_{0,2}$ states. In the case (NR), the net decay rate of the
trapped atom is given as $a_{\rm NR}=a_0+a_2=6.55~$s$^{-1}$,
simply from Eqs.~(3,5). In the case (A),  however, the atoms in
the $^3P_2$ state are distributed to the $^3P_0$ and $^3P_2$ states
and $a_0$ in Eq.~\eqref{eq3} becomes
$a_0'=a_0+\frac{\lambda_{1}}{\lambda_{1}+\lambda_{0}}a_2 = 6.28$,
from the $^{1}S_{1}$
state to the $^{3}P_{0,1,2}$ states. So, the net decay rate of
the trapped atoms for the case (A) is $a_{\rm A}= 6.28$~s$^{-1}$.
Likewise, in the case (B), $a_2$ in Eq.~\eqref{eq5} becomes
$a_2'=a_2+\frac{\lambda_{2}}{\lambda_{2}+\lambda_{1}}a_0 = 3.94$
and, therefore, the decay rate for the case (B) is obtained as
$a_{\rm B}=3.94$~s$^{-1}$. In the case (A+B), the both
$^{3}P_{0,2}$ states are repumped, and the net decay rate to the triplet $P$
states become zero, (i.e., $a_{\rm A+B}=0$).

\section{Results and Discussions}
\begin{figure}[tb]
\centering
\includegraphics[width=0.50\textwidth]{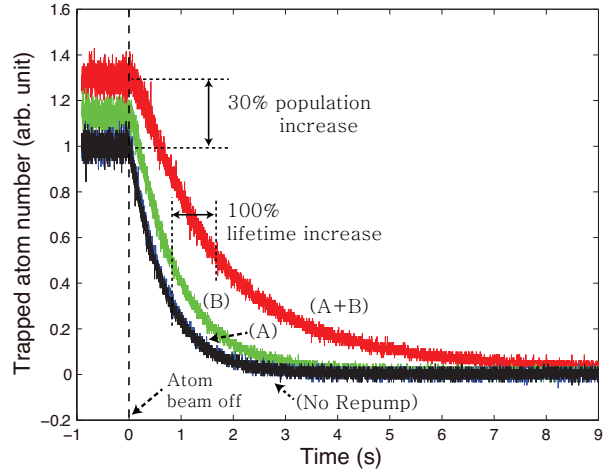}
\caption{ The decay dynamics of $^{174}$Yb MOT for the four
different repumping cases: (A) The $^3P_2$ state, (B) $^3P_0$,
and (A+B) both the states are repumped. The trapped atom
numbers are estimated with the fluorescence from the trapped atoms
in the $^1P_1$ state. The experimental data for the case (A) is
almost overlayed with the (No repump) case.} \label{fig5}
\end{figure}

We investigate the decay dynamics of the trapped atoms in the MOT by shutting off the atomic beam and measuring
the $^1P_1$-$^1S_0$ fluorescence signal as a function of time. 
The experimental situation for zero Yb flux is the condition $\eta=0$ in Eq.~\eqref{eq1}, and its solution is given by
\begin{eqnarray}
N(t)=N(0){e^{-\Gamma t}}\left[1+\frac{\beta N(0)}{\Gamma}\Big(1-e^{-\Gamma t}\Big)\right]^{-1}\;, \label{eq7}
\end{eqnarray}
where $\Gamma$ is the loss rate of the MOT defined as similarly as the parenthesis in Eq.~\eqref{eq1} as
\begin{equation}
\Gamma=a_{x}f(P_T,\Delta)+\gamma_c(f)\;, \label{eq8}
\end{equation}
and the index $x$ in $a_x$ indicates a particular experiment among the four repumping cases (i.e. $x \in \{(NR), (A), (B), (A+B)\}$). 
Figure~\ref{fig5} shows a typical behavior of the temporal
evolution of the number of trapped atoms $N(t)$ in the MOT, for the all four repumping cases. Both the lifetime
$\tau$ of the MOT, which is measured as the $1/e$ decay time, and the steady-state
trapped atom number $N(0)$ are increased by the repumping of the metastable states.
The figure shows that, when the repuming scheme (A+B) has increased $\tau$ by 100\%,  
$N(0)$ has increased by only 30\%.
This result is not consistent with the fact that the change of $N(0)$ should be proportional to the change of $\tau$, or $N(0)=\eta \tau$, 
as $\tau=1/\Gamma$ and $N_{0}={\eta}/{\Gamma}$ from Eq.~\eqref{eq1} and Eq.~\eqref{eq8}.
However, the collisional loss term $\gamma_c$ in Eq.~\eqref{eq8} is,
in fact, factored into two parts: one due to the background
residual gas in the chamber, and the other due to the Yb flux from
the Zeeman slower. Therefore, the collisional
loss term is a function of the loading rate $\eta$, i.e.
$\Gamma=\Gamma(\eta)$, and, as a result, shutting off the atomic beam has suddenly reduced the
loss rate $\Gamma$.
It is noted that the last term in
Eq.~\eqref{eq1} is responsible for the collision between cold Yb
atoms captured in the MOT, not the collision between the Yb in the
MOT and the Yb in the atomic beam from the Zeeman slower.

\begin{figure}[tb]
\centering
\includegraphics[width=0.50\textwidth]{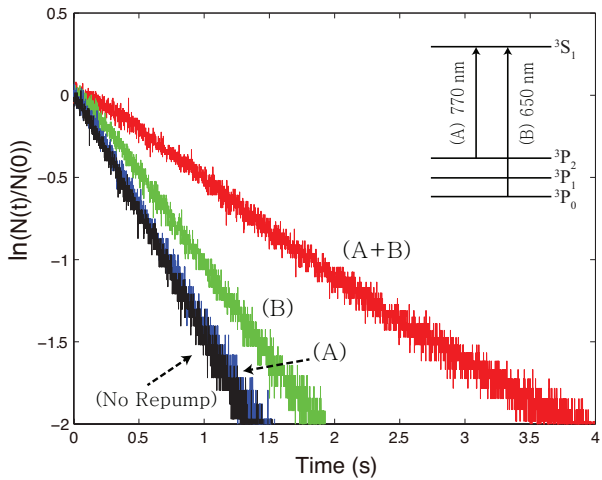}
\caption{Measurement of the temporal evolution of the trapped atom number of the $^{174}$Yb MOT. The result is induced from the $^1P_1$-$^1S_0$ fluorescence signal obtained for the four differenct repumping cases: (NR) no repumping, (A) repumping $^3P_2$ state only, (B) repumping $^3P_0$ state only, and (A+B) repumping the both states. The inset shows the corresponding $^{174}$Yb energy levels.} \label{fig3}
\end{figure}

Figure~\ref{fig3} shows the measured results in a logarithmic scale, 
where $N(t)$, scaled with $N(0)$,
is plotted as a function of time.
It is noted that ${\beta N(0)}/{\Gamma}$ varies from 0 to 1 within the given experimental condition,
so the second term in the paranthesis in Eq.~\eqref{eq7} needs to be included in a fitting of $\Gamma$.
This is especially clear from the behavior of the (A+B) data, shown in red in Fig.~\ref{fig3},
which deviates significantly from a linear line in the given logarithmic plot.

\begin{figure}[tb]
\centering
\includegraphics[width=0.45\textwidth]{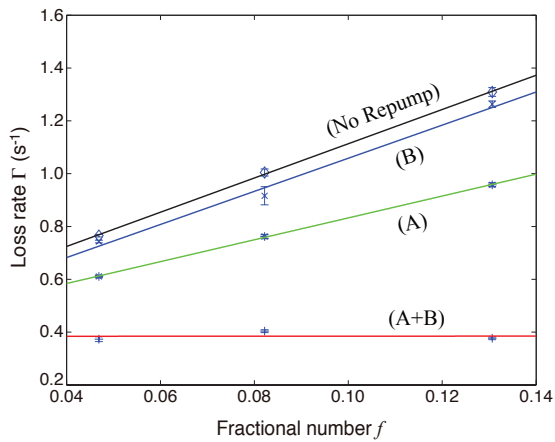}
\caption{Measured loss rate ($\Gamma$) versus the number of atoms in the excited state ($f$), obtained for the four different repumping cases. The slopes ($\Delta \Gamma / \Delta f$) were measured as  6.48, 6.27,  4.14, and  less than $6.3 \times 10^{-3}$, from the top to the bottom curves.} \label{fig4}
\end{figure}

In order to measure $\Gamma$ for the four different
repumping cases, we have performed an experiment for the temporal
evolution of the $^1P_1$-$^1S_0$ fluorescence at three different trap laser power
conditions. The data have been
numerically fitted to Eq.~\eqref{eq7} to obtain $\Gamma(f)$, and the result is shown in Fig.~\ref{fig4}.  
The graphs show the linear dependence of
$\Gamma$ on $f$ for the all four repumping cases. 
Linear regression
analysis predicts that all the four lines can be extrapolated to
converge at $0.42\pm 0.03$ as $f$ approaches zero. It is also
predicted that the slope of the fitted lines, that is the decay
rate $a_x$ in Eq.~\eqref{eq7}, turns out that $a_{\rm  NR}$=6.48,
$a_{\rm A}$=6.27,  $a_{\rm B}$=4.14, and  $a_{\rm A+B} \le 6.3
\times 10^{-3}$. It is noted that the slope for the (A+B) case
suggests that the back-ground collision rate of the trapped atom
is little dependent on $f$.

Using the obtained loss rates, we can estimate the decay rate $a_0$, $a_2$, and $a_{0,2}$, etc, and the result is summarized in Table~\ref{table1}.
Arbeit the measurement uncertainty, the newly measured values for the individual decay rate $a_0$ and $a_2$ show a good agreement with the theoretically predicted values, and the measurement accuracy is significantly improved, compared to the previous experiments carried out by Loftus et al.~\cite{Loftus} The main uncertainty comes from the estimation of $f$, as the power fluctuation and spatial beam profile, and the measurement error of the trap laser causes the uncertainty of the saturation parameter $s$. Also, the uncertainty of  $\Delta/\Gamma$ is caused by the magnetic field in the trap, laser frequency detuning, etc, and its uncertainty is estimated by 30\%. Including the uncertainty in error fitting, the resulting uncertainty of the decay rate is estimated to 33\%. As shown in Table~\ref{table1}, within the range of uncertainty our measurement agrees well with the theoretically predicted results.
\begin{table}[h]
\caption{\label{table1}
Summary of the measured decay rates $a_0$ and $a_2$ of $^{174}$Yb atoms. Numbers are all in inverse second.}
\begin{tabular*}{\hsize}{@{\extracolsep{\fill}}cccc}
\hline\hline
            &    This work & Previous works~\cite{Loftus, Honda}            & Theory~\cite{Porsev}\\
\hline
$a_{2,0}$   & 6.48 (2.11)  & 23 (11)      &   6.6     \\
$a_{0}$     & 5.96 (1.97)   &       &   6.18     \\
$a_{2}$     & 0.42 (0.14)   &       &   0.37     \\
$a_{1}$         &           & 21.3 (2.6)  &     5.2      \\
\hline\hline
\end{tabular*}
\end{table}

\section{CONCLUSIONS}

We have investigated the trapping dynamics of the $^{174}$Yb MOT by
eliminating the shelving loss to the metastable $^3P_{0,2}$ states via
optical repumping.  At the zero Yb loading limit, the individual decay rate to each metastable state is accurately measured, 
showing an excellent agreement with 
the result of master equation calculation. 
With the use of the  optical repumping,  a faster and a more efficient operation of Yb trapping has become possible, 
and it is hoped that this result may contribute to the BEC research as well as the uncertainty evalutions with opticall lattice clock.

This research was supported by Basic Science Research and Mid-career
Researcher Programs through the National Research Foundation of
Korea (NRF) funded by the Ministry of Education, Science and
Technology (2009-0090843, 2010-0013899).

\end{document}